# Crosstalk Analysis in Quantum Networks: Detection and Localization Insights with photon counting OTDR

A. Rahmouni [1, *], P. Shrestha[1], Y.S. Li-Baboud[1], A.M. Richards[2], Y. Shi[1], M. Merzouki[1], L. Ma[1], A. Migdall[1,3]
A. Battou[1], O. Slattery[1] and T. Gerrits[1, *]

[1]*National Institute of Standards and Technology (NIST), 100 Bureau Drive, Gaithersburg, MD 20899, USA*
[2]*Laboratory for Telecommunication Sciences (LTS), 8080 Greenmead Dr, College Park, MD 20740, USA*
[3]*Joint Quantum Institute, University of Maryland, College Park USA*

[*] *anouar.rahmouni@nist.gov*, *thomas.gerrits@nist.gov*

**Abstract:**

Optical crosstalk from sub-milliwatt classical-channel power into quantum channels presents a significant challenge in quantum network development, introducing substantial noise that limits the network's performance, scalability, and fidelity. Crosstalk can occur between fibers at multiple points, such as connector interfaces at patch panels, within optical devices due to inadequate optical isolation, and in other fiber-related components. Crosstalk also occurs between channels with different frequencies within a single optical fiber, as Raman scattering converts classical light to photons at frequencies that may interfere with the quantum signal. While Raman scattering in optical fibers has been widely studied [1-7], crosstalk between fibers in quantum networks has not received sufficient attention. Here we report a demonstration using photon-counting optical time-domain reflectometry ($\nu$-OTDR) to precisely identify and localize crosstalk between separate channels within the same fiber and between separate fibers. The coexistence of classical and quantum signals in the same network necessitates the use of optical switches for efficient routing and control. Crosstalk characterization of an optical switch reveals that crosstalk depends strongly on cross-connect configuration, with higher levels observed when connections are presumed to be physically closer and lower levels when further apart. Additionally, we found that crosstalk exhibits a pronounced wavelength dependence, increasing over tenfold at longer wavelengths. These findings demonstrate the value of $\nu$-OTDR in diagnosing and mitigating crosstalk in quantum networks. They highlight the importance of optimizing optical switch configurations and wavelength management to minimize noise, ultimately enhancing the scalability, fidelity, and overall performance of quantum networks. This work establishes a foundational approach to addressing crosstalk, paving the way for more robust and efficient quantum network designs.

## 1. Introduction

The development of quantum networks [8, 9] marks a revolutionary step in communication technologies, enabling communication security unattainable through classical signals [10], and the interconnection of quantum computers [11, 12], clusters of quantum sensors [13, 14], and associated devices on local, regional, and global scales. These networks exploit entanglement, such as polarization or time-bin entanglement, for transmitting quantum signals [2, 15, 16]. To achieve these objectives, leveraging the pre-existing infrastructure of classical telecommunication, specifically optical fibers and their associated devices, has become a cornerstone of practical quantum network implementation. For instance, the Washington Metropolitan Quantum Network Research Consortium (DC-QNet) aims to develop a quantum network testbed in the Washington DC area employing pre-existing infrastructure of classical communications [17].

Optical fiber technology is the backbone of classical communication network systems due to its low-loss properties and scalability [18]. The integration of innovations such as optical add-drop multiplexers (OADMs) including dense Wavelength Division Multiplexing (DWDM) and Coarse Wavelength Division Multiplexing (CWDM) [19, 20], Micro-Electro-Mechanical Systems (MEMS) including fiber optical cross-connect switches (FOXS) and routers [21, 22], and other related optical components have revolutionized classical networks. However, adapting these technologies to support quantum communication introduces unique challenges. At the forefront of these challenges is the coexistence of classical and quantum signals in the same infrastructure network, which can lead to significant crosstalk and noise, particularly in optical devices, fiber connectors, and other components within the deployed infrastructure.

Crosstalk arises when classical signals interfere with quantum channels, degrading the performance and reliability of quantum networks. Crosstalk can occur between channels at various points within deployed fiber infrastructure at patch panels, connectors, FOXS, OADMs, filters, and related components, or from Raman scattering noise when the classical and quantum signals coexist in the same fiber. While Raman scattering in optical fibers has been widely studied [1-7], quantum network metrology to quantify and localize crosstalk sources between individual channels will be needed. For instance, off-the-shelf classical devices such as OADMs, FOXS, and connectors typically provide isolation in the range of 30 dB to100 dB [23-25], which is sufficient for classical communication. However, in quantum networks, where noise levels are ideally bounded by the detector dark counts, isolation levels > 100 dB will be needed. For example, 1 dBm from a typical classical signal requires classical-quantum channel isolation exceeding 140 dB to reduce noise to an acceptable level below 100 photons per second.

Conventional optical time-domain reflectometry (OTDR) is one of the most widely used techniques for non-destructive characterization of optical fiber links [26]. ν-OTDR, which employs single-photon detectors such as the superconducting nanowire single-photon detector (SNSPD), has been demonstrated [27]. This technique offers a robust approach for characterizing long optical channel losses in fibers and optical components, making it particularly suitable for *in-situ* and non-destructive characterization of quantum networks [28]. Here, we demonstrate employing ν-OTDR to detect crosstalk in deployed fiber networks (referred to as ν-OTDR crosstalk). ν-OTDR crosstalk is conceptually similar to conventional photon-counting OTDR, but focuses on observing crosstalk at faint light levels transmission at over 100 dB attenuation level between channels which is not possible with traditional photodiodes. This technique enables the detection of leakage from classical signal channels into quantum channels at the single-photon level.

In this work, we introduce the use of ν-OTDR crosstalk to characterize crosstalk and identify its locations within deployed optical infrastructure. ν-OTDR crosstalk is a highly sensitive technique that leverages photon counting to precisely map the distribution of crosstalk across fiber networks. Applied to the NIST Gaithersburg quantum network, as part of DC-QNet [17], this approach allows us to pinpoint specific points of interference, such as optical devices and fiber connections, that contribute to crosstalk, thereby facilitating its mitigation.

In addition to ν-OTDR crosstalk measurements, further crosstalk behavior characterization was performed on a low-loss transparent optical switch device employs piezoelectric beam-steering mechanisms to align light beams between input and output fibers without converting them to electrical signals [29]. Optical switches are critical for managing, controlling, and routing coexistent classical and quantum signals within the same network. Our results reveal that crosstalk depends on the cross-connection configuration of the channels. Specifically, for our test device, the crosstalk is higher when cross-connections are physically closer and lower when they are further apart within the device. Furthermore, we observe wavelength-dependent crosstalk: shorter wavelengths exhibit lower crosstalk, while longer wavelengths result in significantly higher crosstalk levels.

These findings underscore the significant role of ν-OTDR in diagnosing and mitigating crosstalk within quantum networks, introducing a novel diagnostic capability. Furthermore, the results emphasize the importance of optimizing optical cross-connections in quantum networks and implementing wavelength management strategies to significantly reduce crosstalk. These advancements are essential for enhancing the scalability, fidelity, and overall reliability of quantum networks. This work establishes a foundational methodology for addressing crosstalk, representing a critical step toward the development of more robust, efficient, and scalable quantum communication systems.

## 2. Crosstalk in deployed fiber.

Connecting a nominally dark fiber to an SNSPD can result in a significant background noise count rate. Without filtering, we have observed noise levels reaching up to $10^5$ counts per second in some dark fibers within our metropolitan DC-QNet network. Such noise can severely disrupt quantum signals if the fiber is intended for quantum network communication. Utilizing wavelength-selective filters, such as DWDMs or CWDMs, can reduce the count rate to $< 10^3$ counts per second, as observed in our network. However, this reduction is limited by noise originating from photon leakage from classical communication signals, typically aligned with ITU grid wavelengths. If the wavelength of the quantum signal overlaps with the leaked classical photons, filtering becomes ineffective. For instance, at the NIST Gaithersburg campus, we observed this problem when classical communication signals leaked into dark fibers. In the experiment setup, as shown in Figure 1a, we employed a tunable filter and a SNSPD to measure the single-photon spectrum in a nominally dark deployed fiber over the O-band wavelength range. The results, presented in Figure 1b, revealed four prominent peaks at O-band. These peaks suggest the presence of photons originating from classical communication lasers in neighboring optical fibers. The observation of these peaks provides clear evidence of crosstalk between fibers, one carrying classical the other quantum.

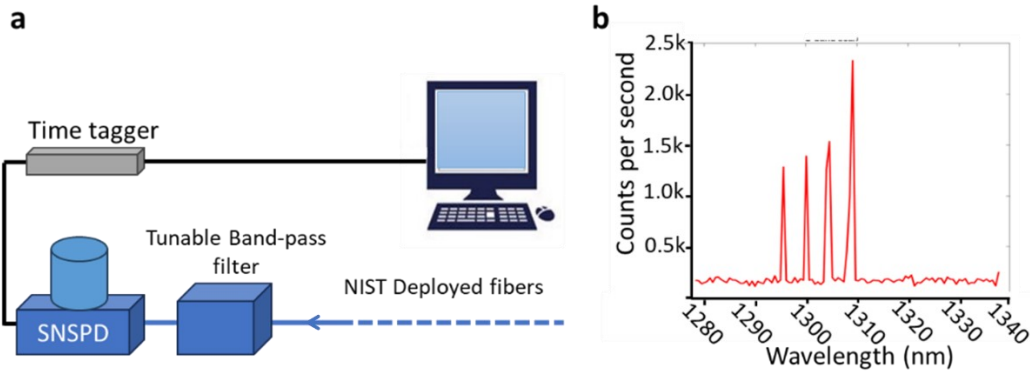

**Figure 1:** Spectral measurement of noise in a dark deployed fiber. a) Optical setup: a nominally dark fiber is connected to a tunable filter, which is subsequently connected to a superconducting nanowire single-photon detector (SNSPD). The count rate is measured as a function of wavelength to obtain the noise spectrum, with the tunable filter operating in the O-band. b) Measured single-photon spectrum: Three distinct peaks from neighboring classical signals we observed within the O-band.

Such similar crosstalk can occur in optical devices such as optical switches, which have crosstalk isolation in the order of 50 dB. Most DWDMs and CWDMs have crosstalk isolation less than 30 dB, while multi-fiber push-on (MPO) connectors can exhibit crosstalk below 100 dB between individual fibers.

## 3. ν-OTDR crosstalk measurement.

We employed ν-OTDR crosstalk to characterize and pinpoint the precise location of the previously mentioned crosstalk (Figure 1b) observed in deployed fiber networks. In our optical setup (Figure 2.a), a supercontinuum laser was used to generate pulsed signals, and a tunable filter selected a wavelength of 1550 nm. This setup measured crosstalk between two deployed nominally dark fibers co-located within the NIST Gaithersburg campus network. In the experiment, a weak optical signal (microwatt level) at 1550 nm was transmitted through one dark fiber (not connected to any active devices) while a second dark fiber located in the same fiber bundle was connected to the SNSPD for photon detection. The laser operated at a repetition rate of 1 kHz and 100 ps pulse width, and the SNSPD was used with a detection efficiency of 85%. Both the laser pulse triggers and photon detection events were recorded using a time tagger, enabling time-of-flight measurements to determine the locations of crosstalk. The detected photon count rate also provided a quantitative estimate of crosstalk intensity. As shown in Figure 2b, the ν-OTDR crosstalk trace reveals several peaks, corresponding to multiple crosstalk locations along the deployed fibers. By converting the time-of-flight data into distances, we identified the specific locations of crosstalk relative to the source. Comparing these locations with the deployed fiber mapping of the NIST Gaithersburg infrastructure revealed that MPO connectors were the sources of the crosstalk peaks observed in Figure 1. The inset of Figure 2b illustrates a schematic of MPO connectors [30], which can support 8, 12, 24, or 48 fibers depending on the application. The typical distance between fibers in an MPO connector is 0.25 mm, resulting in a crosstalk level higher -100 dB.

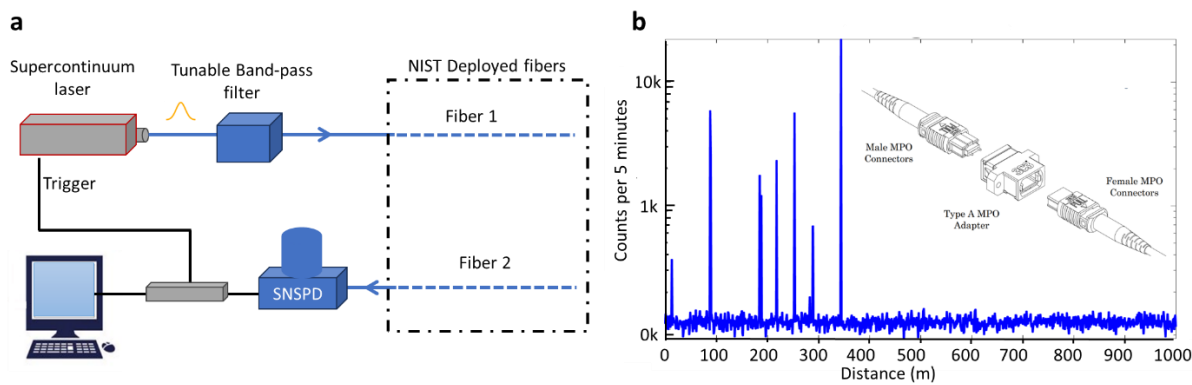

**Figure 2:** Single-photon OTDR crosstalk measurement. a) Optical setup: A classical pulsed signal is generated using a supercontinuum laser, and a tunable filter is used to select a wavelength of 1550 nm. Faint light is sent through a dark fiber, while a nominally dark fiber that is physically adjacent (in the MPO connector) to the illuminated fiber is connected to the SNSPD to detect light leakage from the classical fiber into the dark fiber. The SNSPD detects photons, and both the trigger signal and SNSPD output are connected to a time tagger to measure the time difference between the trigger and detection events, indicating the time-of-flight of photons. This time-of-flight measurement enables the localization of crosstalk within the dark fiber. b) Histogram of the single-photon OTDR measurement: The observed peaks correspond to crosstalk at multiple locations. By converting the time-of-flight data into distance relative to the source, the specific locations of crosstalk peaks are identified. Inset: Illustration of the MPO connector, identified as the source of the crosstalk.

### 4. Crosstalk in optical switches.

All-optical switches are essential for managing quantum and classical signal traffic in quantum network configurations. However, off-the-shelf optical switch devices can exhibit significant crosstalk due to their internal interconnections, allowing leakage from classical channels (fiber carrying classical signals) to quantum channels (fiber carrying quantum signals). Figure 3 illustrates an optical setup designed to characterize the crosstalk performance of an optical 8×8 switch. This switch features 8 input ports (labeled "1" to "8") and 8 output ports (labeled "9" to "16"), with the ability to cross-connect any input to any output port. To characterize the crosstalk, we configured port 1 as the input for a classical signal and port 9 as the output for a quantum signal. A supercontinuum laser equipped with a tunable filter was used to introduce a 3.5 µW of classical power, with 100ps into the optical switch at a specific wavelength. At the output, an SNSPD was employed to measure the crosstalk by counting photons detected at port 9.

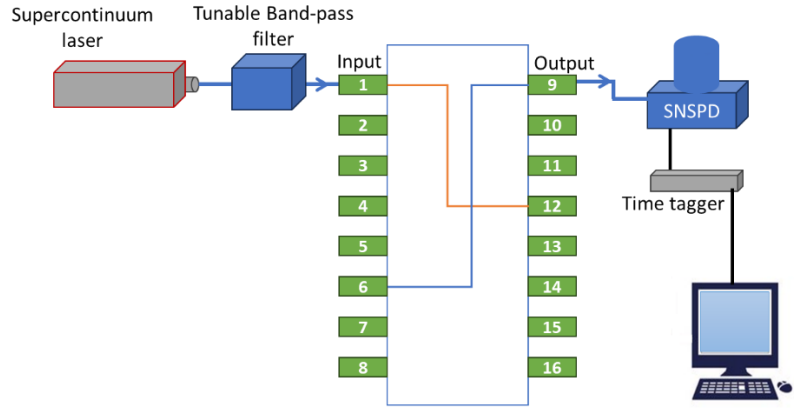

**Figure 3:** Optical setup of crosstalk characterization of an optical switch. The optical 8×8 switch has 8 input ports (numbers "1" to "8"), each of which can be cross connected to any of the 8 output ports (numbers "9" to "16"). The classical optical light connected to port 1. Port 9 connected to SNSPD to detect the photon leaks from a classical signal for different cross connection configuration.

Figure 4(a) presents the crosstalk characterization results obtained by configuring port 1 as the input for a classical signal and port 9 as the output for a quantum signal. Photon counts were measured to quantify the crosstalk from port 1 to port 9 under various cross-connection configurations. The results suggest that crosstalk may be influenced by the physical proximity of cross-connections within the switch, potentially increasing when connections are closer. For instance, when input port 1 is connected to output port 10 and input port 2 is connected to output port 9, the crosstalk is maximized, as shown by the first bar in the graph. In contrast, the crosstalk is substantially lower when the connections are further apart. This observation highlights the importance of carefully selecting cross-connection ports for classical and quantum signals in optical switches to minimize crosstalk, thereby enhancing the performance of quantum networks.

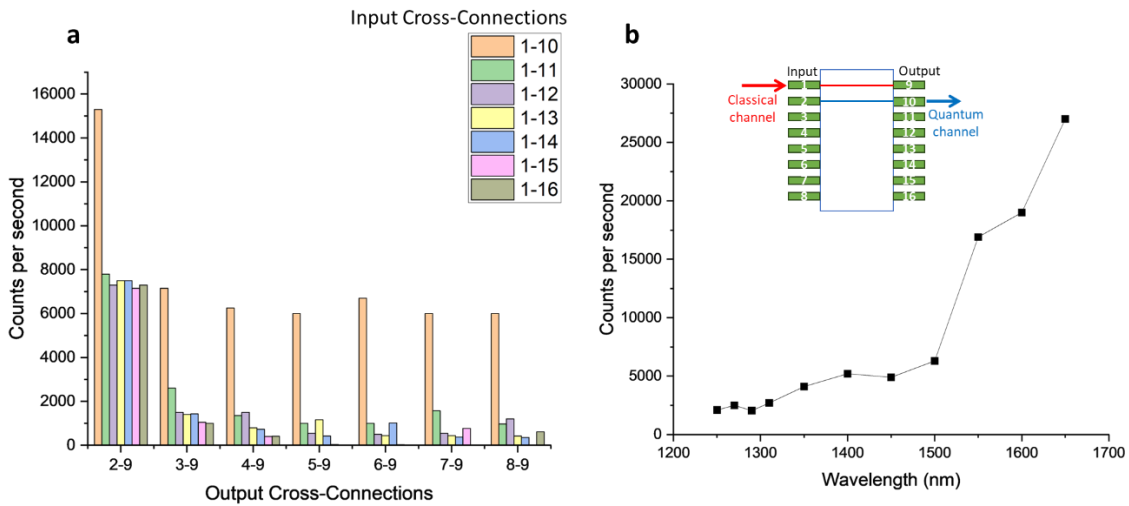

**Figure 4:** Crosstalk characterization of an optical switch with 8 input and 8 output ports. a) Crosstalk from input port 1 to output port 9, measured by varying the cross-connections: connecting input port 1 to any of the 8 output ports (ports 9 to 16) and connecting output port 9 to any of the 8 input ports (ports 2 to 8). b) Wavelength dependence of crosstalk between the cross connection of input port 1 is connected to output port 10 and input port 2 is connected to output port 9. A tunable filter is used at input port 1 to scan the classical wavelength, while input port 2 remains unconnected to any signal.

Characterizing the wavelength dependence of crosstalk is crucial for optimizing wavelength allocation for classical and quantum signals. Figure 4(b) illustrates the measured crosstalk as a function of wavelength, using the cross-connection configuration that exhibited the highest crosstalk in the previous measurements.

In this configuration, input port 1 was connected to output port 9 for the classical signal, while input port 2 was connected to output port 10 for the quantum signal. The results show that crosstalk is significantly lower at shorter wavelengths but increases by an order of magnitude at higher wavelengths. This can be attributed to the extended reach of the evanescent wave at longer wavelengths, which enhances coupling between different cross-connections within the optical switch, leading to increased crosstalk.

5. **Conclusion and discussions:**

Crosstalk noise from classical signals to quantum channels presents a significant challenge in the development of quantum networks. This crosstalk can arise at multiple points within the deployed fiber infrastructure, including optical devices, fiber connections and related components. In this work, we have demonstrated the potential of ν-OTDR crosstalk as a powerful technique for identifying and localizing crosstalk points in deployed fiber networks. By applying this technique to the NIST Gaithersburg infrastructure, we successfully characterized and pinpointed sources of crosstalk, enabling targeted mitigation strategies for noise reduction in our quantum network testbed for the DC-QNet.

Further characterization of optical switches revealed the significant influence of cross-connection configurations on crosstalk behavior. Specifically, we observed that crosstalk is higher when cross-connections are physically adjacent and lower when they are farther apart. Additionally, crosstalk exhibits a strong wavelength dependence: it is significantly lower at shorter wavelengths and increases by an order of magnitude at longer wavelengths. These findings suggest that the O-band is more favorable for classical signals, and C-band is for quantum signals to minimize crosstalk between individual fibers in a quantum network.

While leveraging existing network infrastructure is essential for the practical deployment of quantum networks, significant adaptations are required to meet the stringent demands of quantum communication. Improvements, such as increasing the spacing between fibers to enhance isolation are necessary, increasing the isolation of OADM filters to up to 140 dB (far exceeding the levels achieved in classical networks), and carefully managing cross-connection configurations in optical switches are essential for mitigating crosstalk and interference between channels.

**Disclosures**

The authors declare no conflicts of interest.